\documentclass[aps,prd,onecolumn,eqsecnum,amsmath,nofootinbib,preprintnumbers]{revtex4}%

\usepackage{color,graphicx,float,subfigure,xcolor}%[dvips]%
\usepackage{amsfonts,amssymb,theorem,mathrsfs,times}
\usepackage{bm}
\usepackage{multirow}
\usepackage{mathtools}
\usepackage{amsfonts,amssymb,theorem,mathrsfs}
\usepackage{dsfont}
\usepackage{setspace}
\usepackage{amsmath} 
\usepackage[colorlinks,linkcolor=blue,anchorcolor=blue,citecolor=blue]{hyperref}   % a package related to hyperlink
\usepackage{ulem}   % a package related to strikethrough
\usepackage[english]{babel}

{\theorembodyfont{\upshape}
	}
{\theorembodyfont{\upshape}
	}
{\theorembodyfont{\upshape}
	}
{\theorembodyfont{\upshape}
	}
{\theorembodyfont{\upshape}
	}
{\theorembodyfont{\upshape}
	}
\addtocounter{MaxMatrixCols}{10}
\newcommand{\dalm}{\kern1pt\vbox{\hrule height 0.9pt\hbox{\vrule width
			0.9pt\hskip 2.5pt\vbox{\vskip 5.5pt}\hskip 3pt\vrule width
			0.3pt}\hrule height 0.3pt}\kern1pt}

\begin{document}
\thispagestyle{empty}
% \preprint{\hfill {\small {ICTS-USTC/PCFT-24-03}}}
%<<<<<<<<<<<<< TITLE >>>>>>>>>>>>>>>%
\title{The stability of the greybody factor of Hayward black hole}
	
%<<<<<<<<<<<<< AUTHOR >>>>>>>>>>>>>>>%
%\author{$^b$}
%
% \email{}

% \author{Jia-Ning Chen$^a$\footnote{e-mail address: chenjianing22@mails.ucas.ac.cn}}

\author{Liang-Bi Wu$^{a\, ,c}$\footnote{e-mail address: liangbi@mail.ustc.edu.cn}}

\author{Rong-Gen Cai$^{d\, ,a\, ,b}$\footnote{e-mail address: cairg@itp.ac.cn}} 

\author{Libo Xie$^{a\, ,b\, ,c}$\footnote{e-mail address: xielibo23@mails.ucas.ac.cn (corresponding author)}}

% \author{Li-Ming Cao$^{d\, ,e}$\footnote{e-mail address: caolm@ustc.edu.cn}} 

% \author{Yu-Sen Zhou$^c$\footnote{e-mail
% 		address: zhou\_ys@mail.ustc.edu.cn}}	
	
%<<<<<<<<<<<<< ADDRESS >>>>>>>>>>>>>>>%
\affiliation{${}^a$School of Fundamental Physics and Mathematical Sciences, Hangzhou Institute for Advanced Study, UCAS, Hangzhou 310024, China}

\affiliation{${}^b$CAS Key Laboratory of Theoretical Physics, Institute of Theoretical Physics, Chinese Academy of Sciences, Beijing 100190, China}

\affiliation{${}^c$University of Chinese Academy of Sciences, Beijing 100049, China}

\affiliation{${}^d$Institute of Fundamental Physics and Quantum Technology, Ningbo University, Ningbo, 315211, China}

%<<<<<<<<<<<<< DATE >>>>>>>>>>>>>>>%
\date{\today}
	
%======================================%
%<<<<<<<<<<<<< ABSTRACT >>>>>>>>>>>>>>>%
%======================================%
\begin{abstract}
In this study, we investigate the stability of the greybody factor of Hayward black holes by adding a small bump to the effective potential. Since the greybody factor depends on frequency, we introduce the $\mathcal{G}$-factor and $\mathcal{H}$-factor to quantitatively characterize its stability. We study the stability of the greybody factor within the equal amplitude method and the equal energy method, respectively. Here, the equal amplitude method can be directly imposed by fixing the amplitude of the bump, while the equal energy method requires a physical definition of the energy of the bump with the assistance of hyperboloidal framework. For both methods, when the location of the bump is close to the event horizon of the black hole, and the closer it is to the peak of the original potential, the larger are $\mathcal{G}$-factor and $\mathcal{H}$-factor, and they are bounded by the magnitude of the amplitude or the energy. More importantly, for the equal amplitude method, two factors tend to a specific value as the location of the bump increases. In contrast, for the equal energy method, two factors converge to zero as the location of the bump increases. Notably, the $\mathcal{G}$-factor and the $\mathcal{H}$-factor are insensitive to the regular parameter of Hayward black hole. Therefore, our results indicate that the greybody factor is stable under specific perturbations.
\end{abstract}

\maketitle
	
%======================================%
%<<<<<<<<<<< Introduction >>>>>>>>>>>>>%
%======================================%	
\section{Introduction}
The ringdown phase of a black hole serves as a crucial probe for testing gravitational theories, and it can be described using the linear perturbation theory of black holes. The ringdown spectrum is characterized by a discrete set of quasinormal modes (QNMs) with complex frequencies~\cite{Berti:2009kk,Konoplya:2011qq}, which encode information about the geometric structure of the background spacetime. As the accuracy of gravitational wave detection enhances, research efforts have shifted towards investigating the QNM spectra of black holes immersed in surrounding matter. This shift is motivated by the fact that black holes do not exist in isolation but are inherently embedded within astrophysical environments~\cite{Barausse:2014tra,Cannizzaro:2024yee}. Initial studies~\cite{Nollert:1996rf,Nollert:1998ys} have shown that QNM spectra are sensitive to small perturbations, challenging the assumption that a good approximation of the effective potential would yield minimal deviations in the observed spectra, and many extensive studies have aligned with this conclusion. Current approaches used to study the spectrum instability can be broadly divided into two categories: one involves modifying the effective potential of the black hole~\cite{Daghigh:2020jyk,Qian:2020cnz,Cardoso:2024mrw,Courty:2023rxk,Berti:2022xfj,Cheung:2021bol,Li:2024npg,Yang:2024vor}, the other involves using pseudospectrum analysis~\cite{Jaramillo:2020tuu,Destounis:2023ruj,Jaramillo:2021tmt,Destounis:2021lum,Arean:2023ejh,Boyanov:2023qqf,Cownden:2023dam,Sarkar:2023rhp,Destounis:2023nmb,Boyanov:2022ark,Carballo:2024kbk,Chen:2024mon,Cao:2024oud,Arean:2024afl,Garcia-Farina:2024pdd,Luo:2024dxl,Warnick:2024usx}.

Recently, some authors have suggested that another universal characteristic, namely the black hole greybody factor, represented by $\Gamma_{lm}$,  where $l$ and $m$ refer to different angular modes, could potentially serve as a valuable tool for modeling the amplitude of the ringdown phase~\cite{Oshita:2022pkc,Oshita:2023cjz,Okabayashi:2024qbz}. It means that the ringdown spectral amplitude $|\tilde{h}_{lm}(\omega)|$ can be modeled by the greybody factor. One significant advantage of modeling by using greybody factors lies in the fact that it requires fewer parameters compared to modeling with superposed QNMs, thereby one can effectively avoid the potential overfitting issue and the start time problem in ringdown modeling. Furthermore, studies on the stability of greybody factors can be found in~\cite{Rosato:2024arw,Oshita:2024fzf,Ianniccari:2024ysv}. For Ref.\cite{Rosato:2024arw}, the authors study the stability of the greybody factor against small perturbations of the system, where the small perturbation is encoded by an infinitesimal P\"{o}schl-Teller bump. An integreted quantity is defined to convey the stability of greybody factors [see Eq.(5) therein]. For Ref.\cite{Oshita:2024fzf}, the authors study the stability of the greybody factor against small perturbation of the system, where the small perturbation comes from effective field theory. They compare the perturbed greybody factor with the unperturbed greybody factor by considering their differences that are of the same order as the time-domain waveform [see Fig.3 therein]. For Ref.\cite{Ianniccari:2024ysv}, the authors use the transfer matrix approach of quantum mechanics to give the stability of the greybody factor analytically. These studies indicate that the greybody factors are stable under small perturbations of the system, which is exactly the opposite of the stability of QNMs.

Penrose's singularity theorem~\cite{Penrose:1964wq,Hawking:1970zqf} states that within the classical general relativity (GR) framework, the occurrence of singularities is inevitable, which leads to the breakdown of physical laws~\cite{Hawking:1976ra,Senovilla:1998oua}. The studies of regular black holes have become an important topic in gravitational physics, aimed at exploring ways to solve the singularity problem in classical GR. Therefore, to avoid the conceptual and physical challenges posed by singularities, several models of regular (non-singular) black holes have been proposed~\cite{Bambi:2023try}. The Hayward black hole, which is one important among regular balck holes, is proposed in~\cite{Hayward:2005gi}. 
The metric of Hayward black hole introduces an additional parameter $\ell$ to avoid the central singularity. Its core geometry resembles a de-Sitter like region, implying finite energy density and curvature. The Hayward black hole can be regarded as an effective quantum correction for the Schwarzschild black hole. The effect of quantum gravity will be very significant when $r$ is smaller than the characteristic length scale $\lambda$ with $\lambda$ being smaller than $\ell$. Among the many regular black holes, the Hayward black hole satisfies the requirement of the limiting curvature conjecture~\cite{Frolov:2016pav}. This conjecture requires that, for a practicable fundamental theory, the absolute maximum value of curvature invariant is limited by a certain value, i.e., $|\mathcal{R}| \leq c/\ell^2$. As is well known, many regular black hole models require GR should be coupled with a special non-linear electrodynamics, but recently, \cite{Bueno:2024dgm} has shown that the Hayward black hole can be derived from a pure gravitational theory, though this theory requires the spacetime dimension $d \geq 5$.

Exploring the gravitational waves of the Hayward black hole is one of the important means to study the quantum effects of black holes. As mentioned earlier, the greybody factor is suitable for modeling gravitational waves, so it is necessary to consider the stability of the greybody factor for the Hayward black hole. In addition, although the greybody factor of Hayward black hole has been studied in~\cite{Konoplya:2023ppx}, the stability of the greybody factor of Hayward black holes is still lacking. Studying the stability of the greybody factor of Hayward black holes can serve as a typical example of studying the stability of the greybody factor of regular black holes. It will help us understand the influence of regular parameters on the stability of the greybody factor of black holes. 

It should be emphasized here that when we study the stability of QNM spectra, given that QNM frequencies are only complex numbers, the way to quantitatively characterize their stabilities is relatively simple. To our knowledge, the formula currently used to quantitatively characterize the stability of spectra is $|\omega^{(\epsilon)}-\omega^{(0)}|/|\omega^{(0)}|$, where $\omega^{(\epsilon)}$ denotes the perturbed QNM frequency and $\omega^{(0)}$ denotes the unperturbed QNM frequency. This formula has been applied to a number of references~\cite{Destounis:2023ruj,Cardoso:2024mrw,Cheung:2021bol,Courty:2023rxk,Cao:2024oud}. Alternatively, some authors propose another formula $\lim_{\epsilon\to0}\delta \omega_n/\epsilon$ to quantitatively characterize the sensitivity of QNM frequencies~\cite{Yang:2024vor}. 

But things are not so direct or simple for the greybody factor, since the greybody factor is a function of $\omega\in\mathbb{R}$. There are various ways to evaluate the differences between the original greybody factor and the perturbed greybody factor. Similar to researches on QNM stability, we also study the stability of the greybody factor by adding a small bump to the original effective potential. In this work, to avoid the dependence of the results on different methods of characterizing stability of greybody factor, we introduce the $\mathcal{G}$-factor and $\mathcal{H}$-factor to quantitatively describe the differences in the greybody factor before and after perturbation. The $\mathcal{G}$-factor is based on the $L^p$ integral, and the $\mathcal{H}$-factor is the Hausdorff distance. By comparing the perturbation-induced variations with the original greybody factor, $\mathcal{G}$-factor provides a dimensionless measure that captures the relative impact of perturbations and offers a holistic perspective on the cumulative behavior of the greybody factor by integrating over the entire frequency range. The $L^p$ integral adds flexibility by enabling different weighting schemes for the variations. This method was used in~\cite{Rosato:2024arw} to characterize the stability of the greybody factor for the case $p=1$. The $\mathcal{H}$-factor (Hausdorff distance) is also used to characterize the stability of the greybody factor, which is first introduced in this work. The $\mathcal{H}$-factor provides a comprehensive measure of the dissimilarity between two sets by considering the maximum distance from each point in one set to the nearest point in the other. Unlike the $\mathcal{G}$-factor, which focuses on the cumulative behavior of changes at individual frequency, $\mathcal{H}$-factor emphasizes a global perspective by measuring the overall difference between the sets of greybody factors before and after perturbation. The detailed definitions of two factors can be found in the main text.  

For convenience, we limit the parameter space of the perturbation in two dimensions, namely, the amplitude and position of the perturbation. Additionally, we categorize our approaches to selecting parameters into two types: the first is called equal amplitude method, and the second is called equal energy method. For the equal amplitude method, just as its name implies, we will discuss the impact of the bump's position on the stability of the Hayward black hole's greybody factor by keeping the amplitude unchanged. Based on the hyperboloidal framework~\cite{PanossoMacedo:2023qzp,PanossoMacedo:2024nkw}, the physical size, i.e., energy norm, carried by the bump can be well defined~\cite{Jaramillo:2020tuu,Gasperin:2021kfv}. Based on it, we will discuss the impact of the bump's position on the stability of the greybody factor by keeping the energy carried by the bump unchanged as the equal energy method.

The remains of this paper are organized as follows. In Sec.\ref{sec:The hyperboloidal framework}, we give the hyperboloidal framework of the Hayward black hole and more details can be found in Appendix.\ref{hyperboloidal_coordinate}. In Sec.\ref{sec: greybody}, we show the stability of the greybody factors of Hayward black hole. Sec.\ref{sec: conclusions} provides the conclusions and discussions. In the Appendix.\ref{app: QNMs}, we provide the QNM frequencies of the Hayward black hole.

\section{The hyperboloidal framework of the Hayward black hole}\label{sec:The hyperboloidal framework}
We will start this section with a brief introduction to the Hayward black hole. The Hayward black hole can be derived from Einstein gravity coupled to a non-linear electromagnetic field~\cite{Kumar:2020bqf,Kumar:2020xvu,Fan:2016hvf}. The action is given by~\cite{Fan:2016hvf}
\begin{eqnarray}\label{action}
    I=\frac{1}{16 \pi} \int \mathrm{d}^4 x \sqrt{-g}(R-\mathcal{L}(F))\, ,
\end{eqnarray}
where $F=\mathrm{d}A$ is the field strength of the vector field, $F \equiv F_{\mu\nu}F^{\mu\nu}$ and the Lagrangian density $\mathcal{L}$ is a function of $F$. The covariant equations of motion are
\begin{eqnarray}\label{EOMs}
    G_{\mu \nu}=T_{\mu \nu}, \quad \nabla_\mu(\mathcal{L}_{F} F^{\mu \nu})=0 \, ,
\end{eqnarray}
where $G_{\mu\nu}$ is the Einsteion tensor and $\mathcal{L}_{F}=\partial\mathcal{L}/\partial F$. The Hayward black hole is obtained from the Lagrangian density given by
\begin{eqnarray}\label{Lagrangian_density}
    \mathcal{L}=\frac{4\mu}{\alpha}\frac{(\alpha  F)^{(\mu+3)/4}}{\Big[1+(\alpha F)^{\mu/4}\Big]^2}\, ,
\end{eqnarray}
where $\mu>0$ is a dimensionless constant and $\alpha>0$ has the dimension of length squared. The Hayward black hole is a spherically symmetric spacetime with the metric ansatz
\begin{eqnarray}\label{Hayward_metric}
    \mathrm{d}s^2=-f(r)\mathrm{d}t^2+\frac{\mathrm{d}r^2}{f(r)}+r^2(\mathrm{d}\theta^2+\sin^2\theta\mathrm{d}\phi^2)\, .
\end{eqnarray}
Solving the covariant equations of motion (\ref{EOMs}) with the ansatz (\ref{Hayward_metric}), we can obtain the metric function which is given by~\cite{Fan:2016hvf}
\begin{eqnarray}\label{Hayward_metric0}
    f(r)=1-\frac{2C}{r}-\frac{2Q^3r^{\mu-1}}{\alpha(r^{\mu}+Q^{\mu})}\, ,
\end{eqnarray}
where $C$ and $Q$ are the integration constants. The square of the ﬁeld strength is $F=Q^4/(\alpha r^4)$. For $C=0$ and $\mu=3$, the solution becomes the Hayward black hole~\cite{Hayward:2005gi}. To see it, let $\alpha=M^2\gamma$ and $Q^3=M^3 \gamma$, then the metric function $f(r)$ can be rewritten as
\begin{eqnarray}\label{metric_function}
    f(r)=1-\frac{{2 r^2}/{M^2}}{\gamma+{r^3}/{M^3}} \, .
\end{eqnarray}
For the metric of Hayward black hole, $M$ is known as the mass of the black hole, $\gamma$ is the regular parameter. The event horizon exists within the condition $\gamma \le 32/27$. It is found that such a metric (\ref{Hayward_metric}) describes the geometry of a regular black hole for the metric function $f(r)$ is non-singular at $r=0$. It is evident that when $\gamma=0$, the Hayward black hole returns to the Schwarzschild black hole. When the parameter $\gamma$ is satisfied with $0<\gamma<32/27$, there are two horizons, i.e., the inner horizon $r_{-}$ and the event horizon $r_{+}$, which are solved by equation $f(r)=0$. For the case of $\gamma=32/27$, the Hayward black hole will become extreme, in other words, the inner horizon and the event horizon coincide with each other, i.e., $r_{+}=r_{-}=4M/3$.

For convenience, we will parameterize the metric (\ref{Hayward_metric}) in terms of the horizons $r_{+}$ and $r_{-}$. As defined, $r_{+}$ and $r_{-}$ satisfy $f(r_{+})=0$, $f(r_{-})=0$, respectively. By solving these two equations inversely, the parameters $M$ and $\gamma$ can be expressed in terms of $r_{-}$ and $r_{+}$,
\begin{eqnarray}\label{relation_M_gamma_rp_rm}
    M =\frac{r_{-}^2+r_{-} r_{+}+r_{+}^2}{2 (r_{-}+r_{+})} \, , \quad \gamma=\frac{8 r_{-}^2 r_{+}^2(r_{-}+r_{+})^2}{(r_{-}^2+r_{-} r_{+}+r_{+}^2)^3} \, .
\end{eqnarray}
In order to characterize different Hayward black holes, we introduce the dimensionless parameter $q$ which is defined by $r_{-}={q}^2 r_{+}$. Then the metric function  will be factorized into the form
\begin{eqnarray}\label{metric_function_factorize}
    f(r)=\frac{(r-r_{+})(r-q^2r_{+})\Big[(1+q^2)r+q^2 r_{+}\Big]}{(1+q^2)r^3+q^4r_{+}^3} \, ,
\end{eqnarray}
in which $q$ and $\gamma$ satisfy the following relation
\begin{eqnarray}\label{relation_gamma_q}
    \gamma=\frac{8 q^4 (1 + q^2)^2}{(1+q^2+q^4)^3} \, .
\end{eqnarray}
From the above relation, we have dimensionless parameter $q\in[0,1]$, which corresponds to $\gamma\in[0,32/27]$. They correspond to each other one by one. Due to the factorial decomposition form of the function $f(r)$, it becomes easy for us to compute the tortoise coordinates in terms of $r$, facilitating subsequent calculations. So we refer to $q$ as the regular parameter of Hayward black hole.

Now, we consider the massless scalar field $\Phi(t,r,\theta,\phi)$ propagating in such a Hayward black hole spacetime, which is controlled by the Klein-Gordon equation
\begin{eqnarray}\label{KG_equation}
	\square\Phi=\frac{1}{\sqrt{-g}}(g^{\mu\nu}\sqrt{-g}\Phi_{,\mu})_{,\nu}=0\, ,
\end{eqnarray}
where $g$ denotes the determinant of the metric of the background geometry $g_{\mu\nu}$. Considering a standard separation of variables of the form
\begin{eqnarray}
    \Phi(t,r,\theta,\phi)=\sum_{l,m}\frac{\Psi_{l,m}(t,r)}{r}Y_{l,m}(\theta,\phi)\, ,
\end{eqnarray}
with $Y_{l,m}$ standing for the spherical harmonic function, Eq.(\ref{KG_equation}) is rewritten as 
\begin{eqnarray}\label{wave_equation}
    \Big(\frac{\partial^2}{\partial t^2}-\frac{\partial^2}{\partial r_{\star}^2}+V_l\Big)\Psi_{l,m}=0\, ,
\end{eqnarray}
where the tortoise coordinate $r_{\star}$ and the effective potential $V_l$ are obtained by 
\begin{eqnarray}\label{tortoise_coordinate_and_effective_potential}
	\mathrm{d}r_{\star}=\frac{\mathrm{d}r}{f(r)}\, ,\quad V_l(r)=f(r)\Big[\frac{l(l+1)}{r^2}+\frac{f^{\prime}(r)}{r}\Big]\, ,
\end{eqnarray}
and $l=0,1,2,3,\dots$ is referred to as the angular momentum number. For convenience, we omit $l$ and $m$ for $\Psi_{l,m}$ in what follows. Here, we explicitly write out the tortoise coordinate $r_{\star}(r)$ with $q\neq1$,
\begin{eqnarray}\label{rstar}
    r_{\star}(r)&=&r+\frac{(q^4+q^2+1) q^2r_{+}\ln (r-q^2r_{+})}{q^4+q^2-2}-\frac{(q^4+q^2+1)r_{+}\ln (r-r_{+})}{2 q^4-q^2-1}\nonumber\\
    &&+\frac{(q^4+q^2+1) q^2r_{+} \ln (q^2 r+q^2r_{+}+r)}{2 q^6+7 q^4+7 q^2+2}\, .
\end{eqnarray}

\begin{figure}[htbp]
	\centering
	\includegraphics[width=0.4\textwidth]{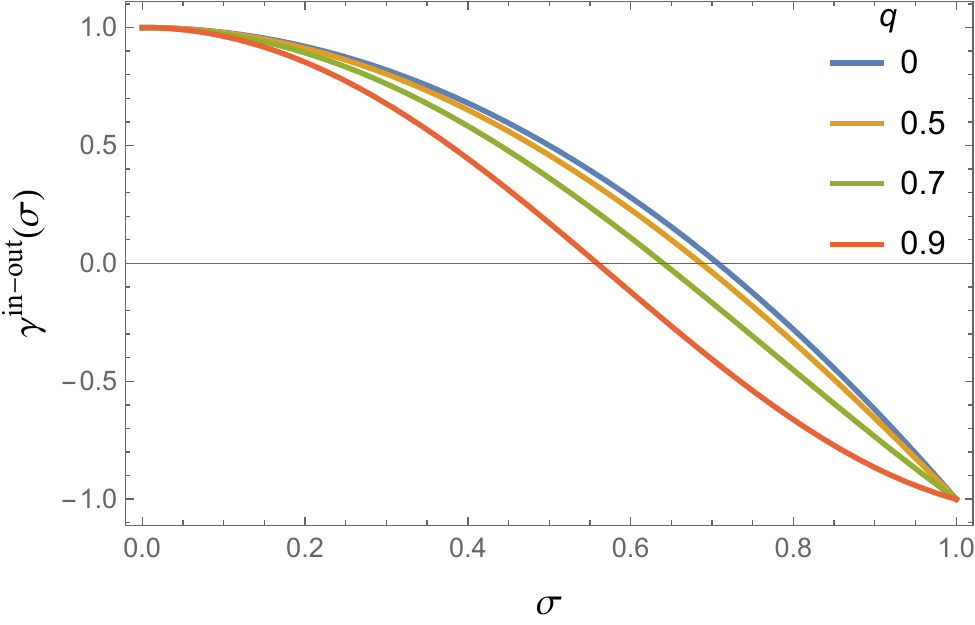}
 \hspace{1cm}
    \includegraphics[width=0.4\textwidth]{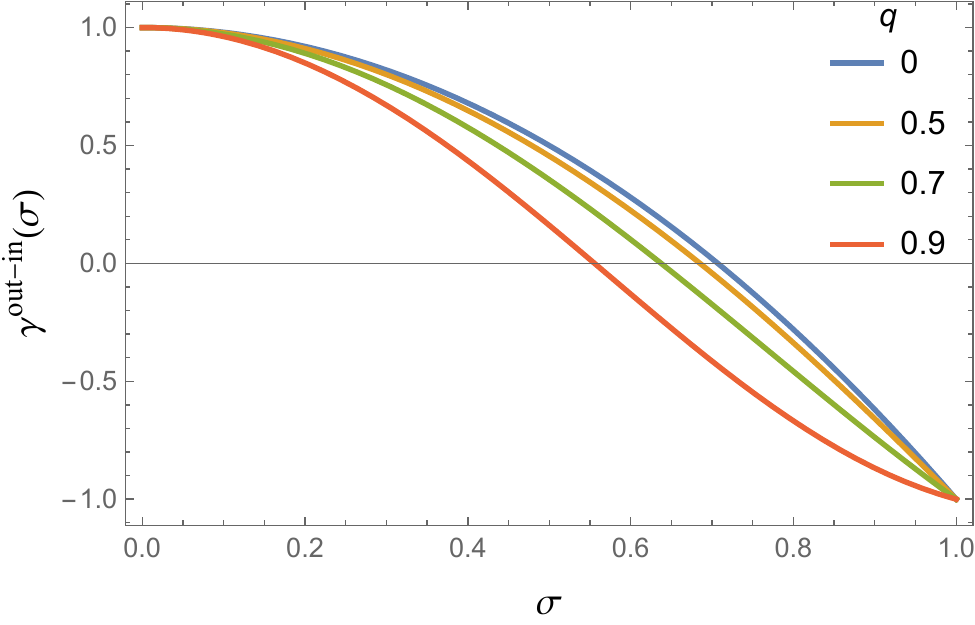}
	\caption{The functions $\gamma(\sigma)$ are depicted  with respect to the coordinate $\sigma$ within different $q$. The left panel stands for the in-out strategy and the right panel stands for the out-in strategy.}
	\label{fig:gamma}
\end{figure}

Next, we will construct the hyperboloidal coordinate transformation and the details have been shown in Appendix.\ref{hyperboloidal_coordinate}. Furthermore, one can refer to~\cite{Zenginoglu:2007jw,PanossoMacedo:2023qzp} to get more information about it. Hence, one gets the resulted partial derivative equation for $\Psi(\tau,\sigma)$ after applying the coordinate transformation (\ref{hyperboloidal_coordinate_transformation}),
% \begin{eqnarray}\label{relationships}
%     \partial_t=\frac{1}{r_{+}}\partial_\tau\, ,\quad \partial_{r_{\star}}=-\frac{\sigma^2\mathcal{F}(\sigma)}{r_{+}}\Big(\partial_\sigma+H^{\prime}(\sigma)\partial_\tau\Big)=-\frac{1}{r_{\star}}\Big(p(\sigma)\partial_\sigma+\gamma(\sigma)\partial_\tau\Big)\, ,
% \end{eqnarray}
\begin{eqnarray}\label{regular_equation}
    w(\sigma)\partial_\tau^2\Psi-2\gamma(\sigma)\partial_\tau\partial_\sigma\Psi-p(\sigma)\partial_\sigma^2\Psi-\partial_\sigma\gamma(\sigma)\partial_\tau\Psi-\partial_\sigma p(\sigma)\partial_\sigma\Psi+q_l(\sigma)\Psi=0\, ,
\end{eqnarray}
where the four functions $p(\sigma)$, $\gamma(\sigma)$, $w(\sigma)$ and $q_l(\sigma)$ are given by
\begin{eqnarray}\label{functions_p_gamma_w_ql}
    p(\sigma)=\sigma^2\mathcal{F}(\sigma)\, ,\quad \gamma(\sigma)=p(\sigma)H^{\prime}(\sigma)\, ,\quad w(\sigma)=\frac{1-\gamma^2(\sigma)}{p(\sigma)}\, ,\quad q_l(\sigma)=\frac{r_{+}^2V_l(\sigma)}{p(\sigma)}\, .
\end{eqnarray}
In Fig.\ref{fig:gamma}, we show the functions $\gamma(\sigma)$ for in-out strategy and out-in strategy within different $q$~\cite{PanossoMacedo:2023qzp}. The messages conveyed to us by this figure is that both strategies are reliable to establish the  the hyperboloidal coordinate transformation in the framework of the Hayward black hole. Given that, in the following chapters, our results will be conveyed under the in-out strategy.

\begin{figure}[htbp]
	\centering
	\includegraphics[width=0.45\textwidth]{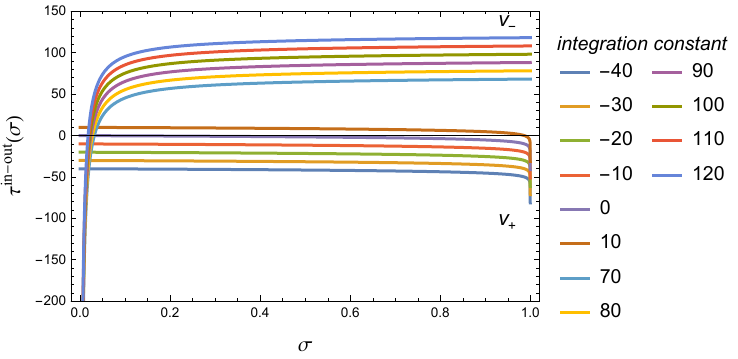}
 \hspace{1cm}
    \includegraphics[width=0.45\textwidth]{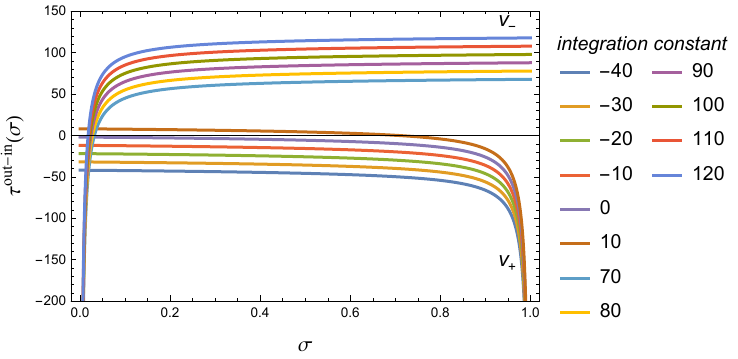}
	\caption{The left panel stands for the in-out strategy and the right panel stands for the out-in strategy. The various colored lines represent the integral curve with different integration constant. Here the curves with integration constants $(-40, -30, -20, -10, 0, 10)$ represent the $v_+$ family, while the curves with integration constants $(70, 80, 90, 100, 110, 120)$ represent the $v_-$ family.} Both of them satisfy the conditions that $v_{+}$ vanishes on the event horizon and $v_{-}$ vanishes on the null infinity.
	\label{fig:characteristic_speed}
\end{figure}

The coefficient functions of the above equation are all regular at the boundaries $\sigma=0$ and $\sigma=1$. Three advantages of hyperboloidal coordinates are abled to be proposed. First, it is imposed outflow boundary conditions on the event horizon $\mathcal{H}^{+}$ and the future null infinity $\mathcal{J}^{+}$, which is consistent with the boundary conditions in terms of the problem of QNMs. Second, the infinite domain $r\in[r_{+},\infty)$ is compactified into a finite domain $\sigma\in[0,1]$. Third, one may extract gravitational waves at the future null infinity $\mathcal{J}^{+}$ which is impossible in traditional Cauchy slices. For the first viewpoint, we can provide the following explanation. The simple reason is based on the hyperbolicity analysis. Two characteristic speeds $v_{\pm}$ are satisfied with the characteristic equation of Eq.(\ref{regular_equation})
\begin{eqnarray}
    w(\sigma)v^2+2\gamma(\sigma)v-p(\sigma)=0\, ,\quad v=\frac{\mathrm{d}\sigma}{\mathrm{d}\tau}\, .
\end{eqnarray}
Therefore, we have characteristic speeds $v_{\pm}$ given by
\begin{eqnarray}
    v_{\pm}(\sigma)=\frac{p(\sigma)}{\mp1+\gamma(\sigma)}\, ,
\end{eqnarray}
where $v_{+}$ is so-called the outgoing characteristic speed and $v_{-}$ is so-called the ingoing characteristic speed. One gets the outgoing characteristic speed $v_{+}$ vanishes on the event horizon $\sigma=1$ and the ingoing characteristic speed $v_{-}$ vanishes on the null infinity $\sigma=0$. Integrating the ODEs $\mathrm{d}\tau(\sigma)/\mathrm{d}\sigma=1/v_{\pm}(\sigma)$, we can get two families of characteristics in Fig.\ref{fig:characteristic_speed}, where the ingoing and outgoing characteristics have been plotted, respectively. This figure tells us that the boundary conditions of QNMs will be automatically embedded in the differential equation (\ref{regular_equation}).

Unlike the Schr\"{o}dinger equation in quantum mechanics, Eq.(\ref{regular_equation}) is second order in time. The simplest choice of reduction into first order in time is to define a new variable as
\begin{eqnarray}
    \Pi=\partial_\tau\Psi\, .
\end{eqnarray}
Then, Eq.\eqref{regular_equation} can be rewritten as a matrix differential equation which involves first order derivative respect to time and second order derivative respect to space, i.e.,
\begin{eqnarray}\label{pde}
    \partial_\tau u=Lu\, ,\quad L=\begin{bmatrix}
        0 & 1\\
        L_1 & L_2
    \end{bmatrix}\, ,\quad u(\tau,\sigma)=\begin{bmatrix}
        \Psi(\tau,\sigma)\\
        \Pi(\tau,\sigma)
    \end{bmatrix}\, .
\end{eqnarray}
Two components $L_1$ and $L_2$ of the operator $L$ are given by
\begin{eqnarray}\label{operator_L_1_L_2}
    L_1=C(\sigma)\partial_\sigma^2+E(\sigma)\partial_\sigma+W(\sigma)\, ,\quad L_2=A(\sigma)\partial_\sigma+B(\sigma)\, ,
\end{eqnarray}
with five functions $C(\sigma)$, $E(\sigma)$, $W(\sigma)$, $A(\sigma)$ and $B(\sigma)$ being expressed as
\begin{eqnarray}\label{functions_C_E_W_A_B}
    C(\sigma)=\frac{p(\sigma)}{w(\sigma)}\, ,\quad E(\sigma)=\frac{\partial_\sigma p(\sigma)}{w(\sigma)}\, ,\quad W(\sigma)=-\frac{q_l(\sigma)}{w(\sigma)}\, ,\quad A(\sigma)=\frac{2\gamma(\sigma)}{w(\sigma)}\, ,\quad B(\sigma)=\frac{\partial_\sigma\gamma(\sigma)}{w(\sigma)}\, .
\end{eqnarray}

So far, we have established the hyperboloidal framework of the Hayward black hole. Here, such hyperboloidal framework refers to the PDE system [Eq.(\ref{pde})-Eq.(\ref{functions_C_E_W_A_B})]. An important application of the hyperboloidal framework is computing QNM frequencies of the Hayward black hole. For the PDE system [Eq.(\ref{pde})-Eq.(\ref{functions_C_E_W_A_B})], performing the Fourier transformation in $\tau$
\begin{eqnarray}
    u(\tau,\sigma)\sim u(\sigma)e^{i\omega\tau}\, ,
\end{eqnarray}
we arrive at the eigenvalue problem as follow
\begin{eqnarray}\label{eigenvalue_problem}
    \frac{1}{i}
    \begin{bmatrix}
        0 & 1\\
        L_1 & L_2
    \end{bmatrix}
    \begin{bmatrix}
        \Psi\\
        \Pi
    \end{bmatrix}=\omega
    \begin{bmatrix}
        \Psi\\
        \Pi
    \end{bmatrix}\, .
\end{eqnarray}
We use the differential matrice $\mathbf{D}$ to discretize the operator $-iL$, and compute the eigenvalues of the matrix $-i\mathbf{L}$, where we choose the Chebyschev-Lobatto grids given by
\begin{eqnarray}
    \sigma_j=\frac{1+z_j}{2}\, ,\quad z_j=\cos\Big(\frac{j\pi}{N}\Big)\, ,\quad j=0,1,\cdots,N-1,N\, .
\end{eqnarray}
Therefore, within the framework of hyperboloidal coordinates, the QNM frequencies can be directly obtained from the eigenvalues of Eq.(\ref{eigenvalue_problem}). We have reproduced QNM frequencies of Hayward black hole in~\cite{Flachi:2012nv,Toshmatov:2015wga} by the pseudo-spectral method, and then we show the QNM frequencies with $q=0$, $q=1/4$, $q=1/2$ and $q=3/4$ in Appendix.\ref{app: QNMs}. Note that we have fixed $M=1$ in the case of obtaining the QNM frequencies.

\section{the stability of greybody factors}\label{sec: greybody}
In this section, we consider the stability of greybody factors of Hayward black hole. Here, we quantitatively calculate the variation of the greybody factor with small perturbations in the effective potential for the Hayward black hole to characterize its stability. Similar to the approach in the reference~\cite{Rosato:2024arw}, we also add a small bump to the original effective potential. But what is important and different is that we will also study the stability of the greybody factor under equal energy perturbations. Simply keeping the amplitude of the bump constant cannot achieve this.

First of all, we give a brief review on the greybody factor. The greybody factor quantifies the absorptive nature of a black hole geometry, and it is determined by the geometry of black holes like the QNM frequenies. We start with the perturbation equation (\ref{wave_equation}) in frequency domain of Hayward black hole, which is described by
\begin{eqnarray}\label{scattering _equation}
    \Big[\frac{\mathrm{d}^2}{\mathrm{d} r_{\star}^2}+\omega^2 -V_l(r)\Big]\tilde{\Psi}(\omega)=0\, ,
\end{eqnarray}
where $\tilde{\Psi}(\omega)$ is the Fourier transformation of $\Psi$. The black hole greybody factor is the transmission coefficient of a scattering problem identified by the following boundary conditions
\begin{eqnarray}\label{boundary_condition}
\tilde{\Psi}(\omega)=\left\{\begin{array}{l}
e^{-i\omega r_{\star}}\, ,\quad r_{\star} \to-\infty \\
A^{\text {in }}(\omega) e^{-i\omega r_{\star}}+A^{\text {out }}(\omega) e^{+i\omega r_{\star}}\, ,\quad r_{\star}\to+\infty
\end{array}\right.\, .
\end{eqnarray}
For the scattering problem with an effective potential, we can define the reflectivity and transmissivity of the background spacetime
\begin{eqnarray}\label{reflectivity_transmissivity}
R_l(\omega)=\Big|\frac{A^{\text{out}}(\omega)}{A^{\text {in}}(\omega)}\Big|^2, \quad \Gamma_l(\omega)=\Big|\frac{1}{A^{\text {in}}(\omega)}\Big|^2\, ,
\end{eqnarray}
in which $\Gamma_l(\omega)$ is precisely the greybody factor and energy conservation enforces $R_l(\omega)+\Gamma_l(\omega)= 1$.

\begin{figure}[htbp]
	\centering
	\includegraphics[width=0.5\textwidth]{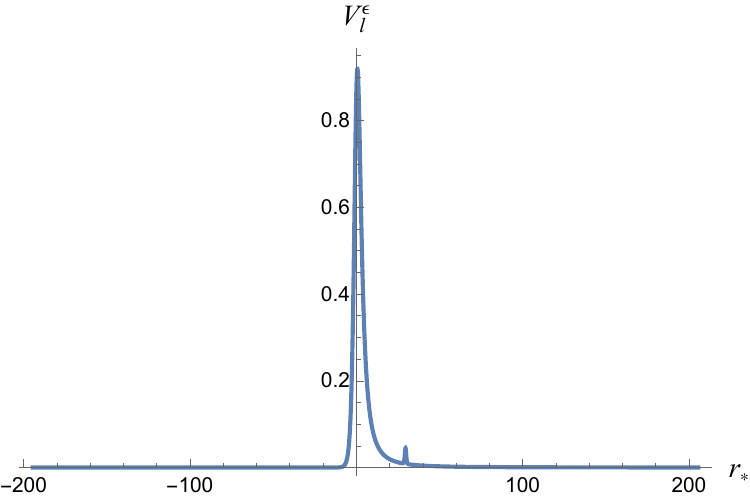}
	\caption{Perturbed effective potential (\ref{perturbed_effective_potential}), for $r_{+}= 1$, $q = 1/2$, $l = 2$ and the perturbation parameters $\epsilon=0.01$, $c_0=26.25$.}
	\label{fig:V_perturbed}
\end{figure}

To study the stability of greybody factors against small perturbations of the system, we consider an infinitesimal P\"{o}schl-Teller bump $V_\epsilon(r)$ added to the original potential, which means that the new potential denoted by $V_{l}^{\epsilon}(r)$ has the form
\begin{eqnarray}\label{perturbed_effective_potential}
V_{l}^{\epsilon}(r)=V_l(r)+V_\epsilon(r)\, ,\quad V_\epsilon(r)=\frac{\epsilon}{M^2} \operatorname{sech}\Big[\frac{r_{\star}(r)-r_{+}c_0}{M}\Big]^2\, ,
\end{eqnarray}
where $V_l(r)$ is the original effective potential of Hayward black hole given by Eq.(\ref{tortoise_coordinate_and_effective_potential}). The addition of such a single small bump intends to simulate some radial concentrated distribution of matter~\cite{Cheung:2021bol,Berti:2022xfj,Cardoso:2024mrw}. In addition, $r_{+}c_0$ and $\epsilon$ respectively parameterize the location and amplitude of the bump, where both $c_0$ and $\epsilon$ are dimensionless. When the effective potential $V_l(r)$ is perturbed to become a new effective potential $V_l^\epsilon(r)$, the corresponding greybody factor will also change, and we denote the new greybody factor as $\Gamma_l^\epsilon(\omega,c_0)$. Given the parameter $r_{+}=1$, $q=1/2$, $l=2$, $\epsilon=0.01$ and $c_0=26.25$, we show the perturbed potential $V_{l}^{\epsilon}$ in terms of the tortoise coordinate $r_{\star}$ in Fig.\ref{fig:V_perturbed}. Numerically, we use $4$-order Runge-Kutta method to solve Eq.(\ref{scattering _equation}) with the boundary condition (\ref{boundary_condition}) to get the greybody factors $\Gamma_l^\epsilon(\omega,c_0)$.

Here, we explain and clarify the range of parameters in numerical calculations. As we know, the fundamental mode $l=m=2$ is the most common mode in astrophysical black holes, which is clearly expected from tensor perturbation and results of numerical relativity, and it occupies a very important position in observations. Therefore, in order to minimize the variation of parameters but better grasp the research of the stability of the greybody factor for the Hayward black hole, our numerical results are based on the angular momentum parameter $l=2$. 

\begin{figure}[htbp]
	\centering
	\includegraphics[width=0.45\textwidth]{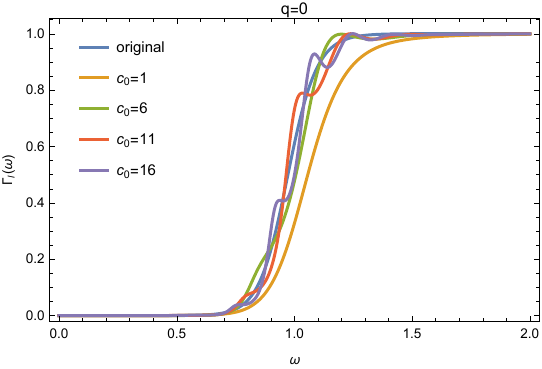}
    \hspace{0.5cm}
    \includegraphics[width=0.45\textwidth]{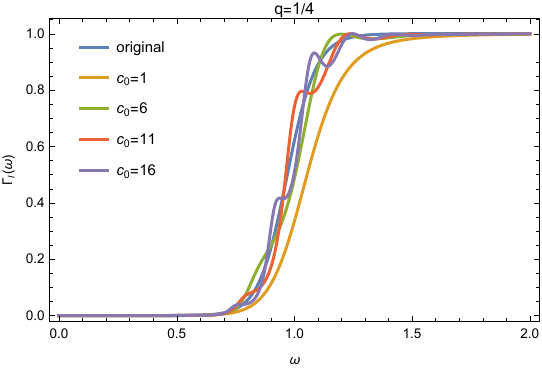}
     \includegraphics[width=0.45\textwidth]{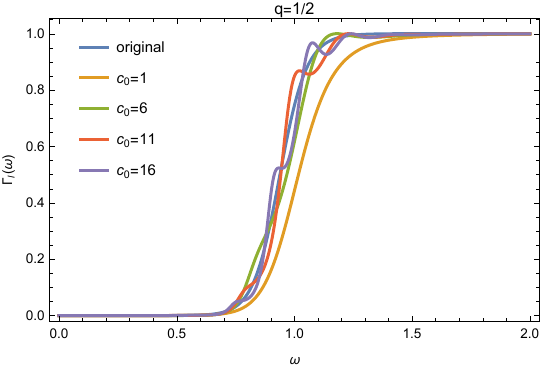}
     \hspace{0.5cm}
    \includegraphics[width=0.45\textwidth]{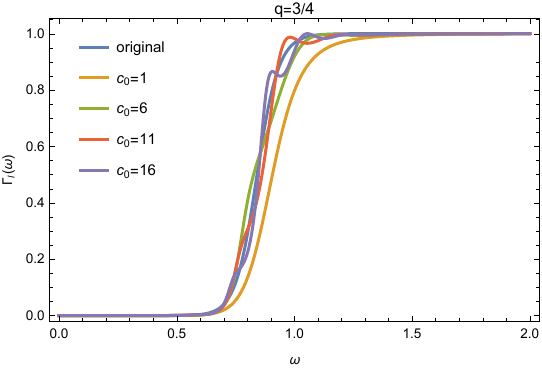}
	\caption{For the the amplitude $\epsilon=1/10$, greybody factors $\Gamma_l(\omega)$ varying with different bump location $c_0$ are depicted in each panels. Four panels correspond to the case of $q=0$, $q=1/4$, $q=1/2$ and $q=3/4$, respectively. The blue curve stands for the original greybody factor for each $q$.}
	\label{fig:Greybody_factor}
\end{figure}

In addition, we notice that we don't need to know the specific value of the scale factor $r_{+}$ in the process of obtaining the QNM frequencies of the non-perturbation equation (\ref{eigenvalue_problem}). However, when we add a bump perturbation in the form of $V_\epsilon(r)$, we have to know the value of $r_{+}$ since $M$ depends on $r_{+}$. Without loss of generality, the scale factor is taken as $r_{+}=1$, and at this time the bump is located at $c_0$. Furthermore, we are more concerned about the situation where the location of this bump is far away from the event horizon. As a result, $c_0$ is limited to be greater than $0$. The larger $c_0$ is, the farther the bump is from the event horizon. In order to amplify the difference between the perturbed greybody factor and the original greybody factor, we take the amplitude $\epsilon=1/10$ and show the greybody factors in Fig.\ref{fig:Greybody_factor} in terms of different $q$.

As the Fig.\ref{fig:Greybody_factor} shows, we present the results of the greybody factors in the frequency domain before and after perturbation with different parameters ($q,c_{0}$). We can find that in the low-frequency and high-frequency regions, the greybody factor is insensitive to the perturbation. The effect of the perturbation on the greybody factor is mainly observed in the intermediate region. Furthermore, by comparing the results, we observe that the location of the bump significantly influences the behavior of the greybody factor. Different $c_{0}$ lead to distinct deviation patterns in the perturbed greybody factors. Roughly speaking, the larger the bump location $c_0$, the more pronounced the oscillation of the greybody factor in the intermediate region. However, the influence of the black hole’s regular parameter $q$ on the greybody factor is not significant. They just quantitatively change the position where the curve oscillates in the frequency domain.

However, the aforementioned statement is merely based on direct visual observations, and we have chosen $\epsilon=1/10$ in order to magnify the observed effect. When $\epsilon$ is very small, it is difficult for us to directly observe and discern differences. Inspired by the above, it is necessary for us to quantitatively characterize the differences or the stabilities. In this study, we propose two ways to evaluate differences, namely providing two definitions for the stability of greybody factors. The first definition of stability which is denoted by $\mathcal{G}_{l}(\epsilon,c_0)$ coming from~\cite{Rosato:2024arw}. To make this method of defining stability through integration more general, we further extend it by adding a parameter $p$ with $p\in\mathbb{Z}^{+}$, i.e., the $L^p$ norm via
\begin{eqnarray}\label{Gfactor}
    \mathcal{G}_{l}(\epsilon,c_0)=\Bigg[\frac{\int_{\omega_i}^{\omega_f}|\Gamma^\epsilon_l(\omega,c_0)-\Gamma_l(\omega)|^p\mathrm{d}\omega}{\int_{\omega_i}^{\omega_f}|\Gamma_l(\omega)|^p\mathrm{d}\omega}\Bigg]^{\frac{1}{p}}\, ,
\end{eqnarray}
in which $\omega_i$ and $\omega_f$ are the upper and lower limits of the integral, respectively. Note that only the case $p=1$ has been studied in~\cite{Rosato:2024arw}. The method of describing the stability of greybody factors through integration can also been found in~\cite{Ianniccari:2024ysv}. $\mathcal{G}_{l}$ can comprehensively evaluate the differences between two functions $\Gamma^\epsilon_l$ and $\Gamma_l$ over the entire interval, not just at a certain point or local area, by calculating the integral value of the $p$-power error of two functions over a specified interval. Since the perturbation we added is a local quantity, both the unperturbed greybody factor and the perturbed greybody factor satisfy $\Gamma\to0$ as $\omega\to0$ and $\Gamma\to1$ as $\omega\to\infty$. Based on the above facts, in numerical setting, $\omega_i$ is a real positive number very close to $0$, and $\omega_f$ is a real positive number which ensures that the perturbed and unperturbed greybody factors are very close to $1$ when both of them are evaluated at $\omega_f$. What needs to be emphasized here is that the upper limit $\omega_f$ of integration of the denominator of Eq.(\ref{Gfactor}) cannot be infinite, otherwise the denominator will be infinite, which will make the whole integral meaningless. As for the integral appeared in Eq.(\ref{Gfactor}), we use the composite trapezoidal formula to obtain such integral. For convenience, we will refer to $\mathcal{G}_{l}(\epsilon,c_0)$ as $\mathcal{G}$-factor below.

The second way of defining the stability of the greybody factor is the (double) Hausdorff distance between the function $\Gamma_l^\epsilon(\omega,c_0)$ and the function $\Gamma_l(\omega)$. The definition of the (double) Hausdorff distance is given as follows. Two sets $A=\{(\omega_1,\Gamma_l(\omega_1)),\cdots,(\omega_s,\Gamma_l(\omega_s))\}$ and $B=\{(\omega_1,\Gamma_l^\epsilon(\omega_1)),\cdots,(\omega_s,\Gamma_l^\epsilon(\omega_s))\}$ are constructed from the numerical results of greybody factors, where $s$ stands for the numeber of points in the set $A$ and $B$. Then, the Hausdorff distance is given by
\begin{eqnarray}\label{Hfactor1}
    \mathcal{H}_l(A,B)=\max(h(A,B),h(B,A))\, ,
\end{eqnarray}
where the definition of $h(A,B)$ is given by
\begin{eqnarray}\label{Hfactor2}
    h(A,B)=\max_{a\in A}\min_{b\in B}|a-b|\, .
\end{eqnarray}
Note that $h(A,B)\neq h(B,A)$ in general and $|\cdot|$ stands for the Euclidean distance. The Hausdorff distance goes beyond merely focusing on a single pair of nearest points or the average distance between sets. Instead, it considers the maximum distance from every point in one set to its nearest point in the other set. This measurement approach allows the Hausdorff distance to capture the overall degree of matching between two sets, rather than being limited to local or average similarities. Similarly, we will refer to $\mathcal{H}_{l}(A,B)$ as $\mathcal{H}$-factor below. Note that in \textit{Mathematica}, one can use the built-in function \textit{RegionHausdorffDistance} to compute it.

In the next two subsections, we will investigate the stabilities of the greybody factor, through the involvements of the $\mathcal{G}$-factor and the $\mathcal{H}$-factor defined above, by using the equal amplitude method and the equal energy method respectively, where the energy (norm) will be defined in the corresponding subsection.

\subsection{The equal amplitude method}
\begin{figure}[htbp]
	\centering
	\includegraphics[width=0.45\textwidth]{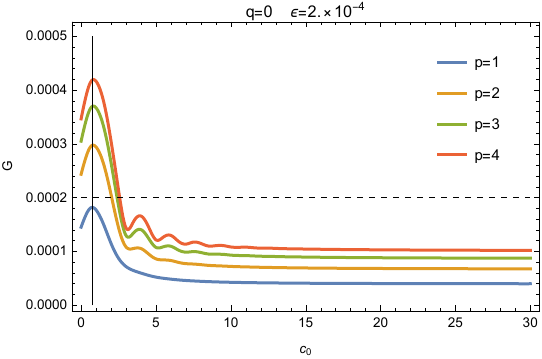}
    \hspace{0.5cm}
    \includegraphics[width=0.45\textwidth]{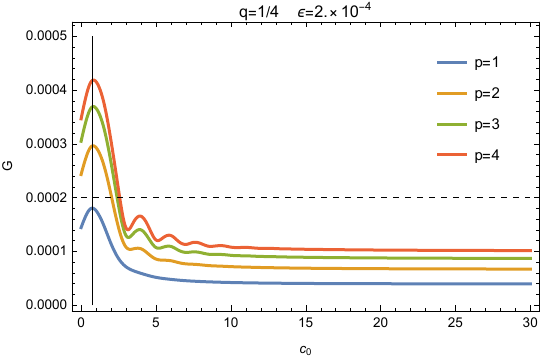}
     \includegraphics[width=0.45\textwidth]{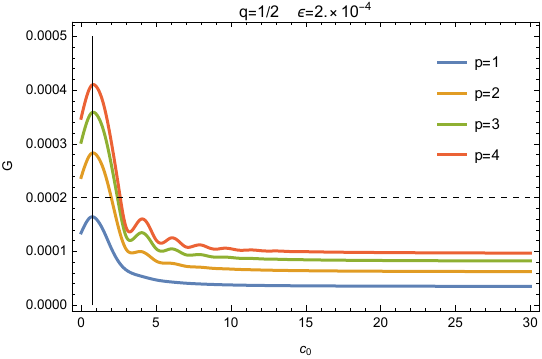}
     \hspace{0.5cm}
    \includegraphics[width=0.45\textwidth]{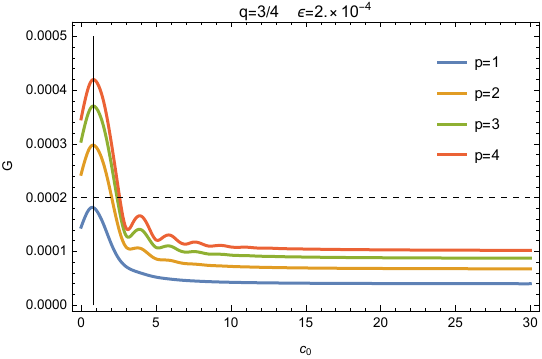}
	\caption{For the bump amplitude  $\epsilon=2\times 10^{-4}$, the $\mathcal{G}$-factor varies with the position of the bump $c_0$ in terms of different $q$ and $p$. Each panel describes the variation of the $\mathcal{G}$-factor changing with $c_0$ and $p$ at a fixed $q$. Here, the vertical black line in each panel stands for the location of the peak of original potential associated with the tortoise coordinate $r_{\star}$ for different $q$. The horizontal dashed line represents the amplitude of the bump we used.}
	\label{fig:G_factor_equal_eps}
\end{figure}

In this subsection, we focus the stability of the greybody factor in the condition of equal amplitude. For the bump amplitude  $\epsilon=2\times 10^{-4}$, we show the $\mathcal{G}$-factor in Fig.\ref{fig:G_factor_equal_eps} within different $q$ associated with the Hayward black hole, respectively. As each panels in Fig.\ref{fig:G_factor_equal_eps} show, when the location $c_0$ of the perturbation bump is close to the event horizon of the Hayward black hole, the $\mathcal{G}$-factor is bounded by the perturbation amplitude $\mathcal{O}(\epsilon)$. The value of $\mathcal{G}_{l}$ is locally maximized when $c_0$ approximately corresponds to the peak of the unperturbed potential, which is consistent with the results in~\cite{Rosato:2024arw}. Our results also show the influence of different norms marked by $p$ on the behavior of the $\mathcal{G}$-factor: as $p$ increases, the instability of the greybody factor of the Hayward black hole (also including the Schwarzschild case) characterized by the $\mathcal{G}$-factor is amplified. But their magnitudes are still around $\mathcal{O}(\epsilon)$. 

\begin{figure}[htbp]
	\centering
	\includegraphics[width=0.45\textwidth]{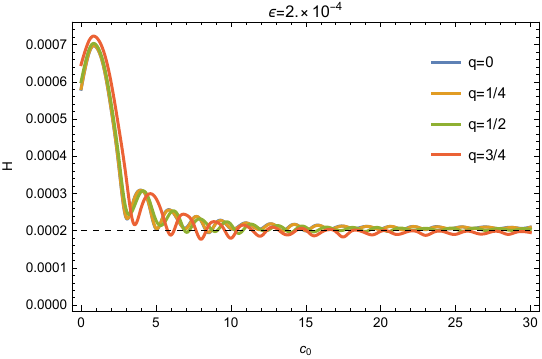}
    \hspace{0.5cm}
    \includegraphics[width=0.45\textwidth]{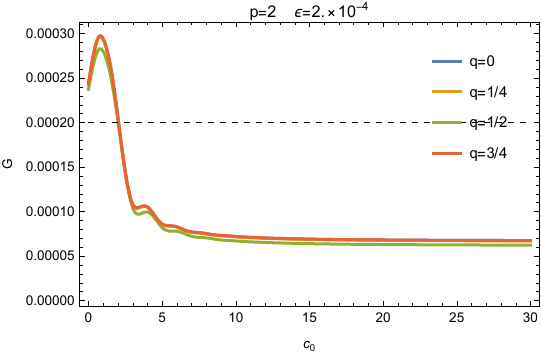}
	\caption{For the bump amplitude  $\epsilon=2\times 10^{-4}$, the $\mathcal{H}$-factor varies with the position of the bump $c_0$ in terms of different $q$ for the left panel. For $p=2$ in Eq.(\ref{Gfactor}), the $\mathcal{G}$-factor varies with the position of the bump $c_0$ in terms of different $q$ for the right panel. The horizontal dashed line stands for the amplitude of the bump we used.}
	\label{fig:H_factor_equal_eps}
\end{figure}

In Fig.\ref{fig:H_factor_equal_eps}, we compare the results of two different methods to characterize the stability of the greybody factor with the condition of equal amplitude ($\epsilon=2\times 10^{-4}$), where different color curves represent different $q$. Note that the right panel of Fig.\ref{fig:H_factor_equal_eps} is inherited from Fig.\ref{fig:G_factor_equal_eps} for convenience of comparison. It can be found that the $\mathcal{G}$-factor and the $\mathcal{H}$-factor are insensitive to the regular parameter $q$ of the Hayward black hole. The most significant difference between them is that the $\mathcal{H}$-factor exhibits oscillatory behavior in region far from the event horizon, while the $\mathcal{G}$-factor does not. However, in terms of the overall trend, the results of two panels show a high degree of consistency, indicating that the greybody factor is a stable physical quantity.

\subsection{The equal energy method}
In the previous subsection, we demonstrated the stability of the greybody factor of the Hayward black hole by adopting the approach of equal amplitude $\epsilon$. To further consolidate our confidence in the stability of the greybody factor, we adopt an alternative method of parameter selection to investigate its stability, which is called the equal energy method (see below). Physically, we use the energy norm defined in~\cite{Jaramillo:2020tuu,Gasperin:2021kfv} to measure the energy (norm) of bump which is denoted by $\lVert V_\epsilon\rVert_E$. The definition of this norm associated with the field $\Psi$ and $\Pi$ in the Hayward balck hole is given by
\begin{eqnarray}\label{energy_norm_sigma}
    \lVert u\rVert_{E}^2&=&\Big\lVert \begin{bmatrix}
        \Psi\\
        \Pi
    \end{bmatrix}\Big\rVert_E^2=\frac12\int_{0}^1\Big[w(\sigma)\Pi^{\star}\Pi+p(\sigma)\partial_\sigma\Psi^{\star}\partial_\sigma\Psi+q_l(\sigma)\Psi^{\star}\Psi\Big]\mathrm{d}\sigma\, .
\end{eqnarray}
Here, the symbol $\star$ stands for the complex conjugate and functions $w(\sigma)$, $p(\sigma)$, $q_l(\sigma)$ are given by Eqs.(\ref{functions_p_gamma_w_ql}), respectively.

Since the above norm is defined in the compact coordinate $\sigma$, we are supposed to rewrite the P\"{o}schl-Teller bump in the the compact coordinate as $V_\epsilon(\sigma)$ by using the coordinate transformations (\ref{hyperboloidal_coordinate_transformation}). Then, as a measurement, we define the energy norm of P\"{o}schl-Teller bump as follows
\begin{eqnarray}\label{Veps_norm}
    \lVert V_\epsilon\rVert_E=\Bigg{\lVert}\begin{bmatrix}
        0 & 0\\
        \frac{r_{+}^2V_{\epsilon}(\sigma)}{p(\sigma)w(\sigma)} & 0
    \end{bmatrix}\Bigg{\rVert}_E=\Bigg{\lVert}W\cdot\begin{bmatrix}
        0 & 0\\
        \frac{r_{+}^2V_{\epsilon}(\sigma)}{p(\sigma)w(\sigma)} & 0
    \end{bmatrix}\cdot W^{-1}\Bigg{\rVert}_2\, ,
\end{eqnarray}
where the first equality represents the size of the P\"{o}schl-Teller bump and the second equality gives us the method to determine such size. At the matrix level, the matrix $W$ comes from the norm (\ref{energy_norm_sigma}) via the Cholesky decomposition of the Gram matrix, in which the Gram matrix derived from the discrete version inner product associated with the norm. Note that $\lVert\cdot\rVert_2$ refers to the $2$-norm and the notion $r_{+}^2V_\epsilon(\sigma)/(p(\sigma)w(\sigma))$ in the third expression becomes a diagonal matrix. One can prove that the limit of $r_{+}^2V_\epsilon(\sigma)/(p(\sigma)w(\sigma))$ on $\sigma=0$ and $\sigma=1$ are both vanished as $q<1$. Regarding more detailed numerical techniques, one can refer to our previous work~\cite{Cao:2024oud}. 

\begin{figure}[htbp]
	\centering
	\includegraphics[width=0.6\textwidth]{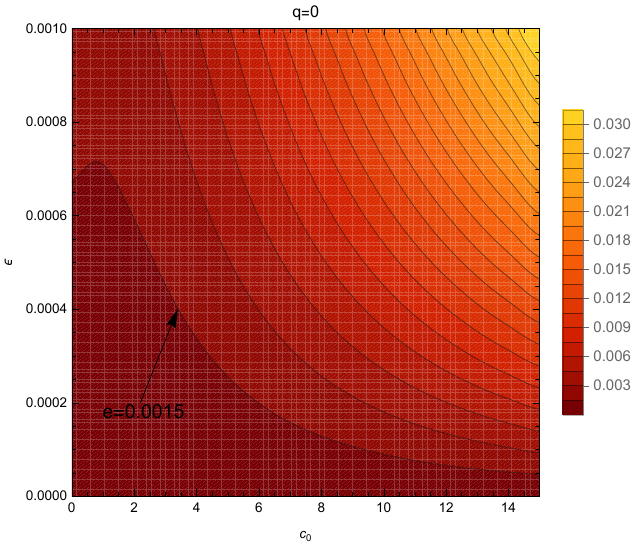}
	\caption{The energy norm contour lines of the bump for the case $q=0$ are shown, in which we specifically point out the energy contour line $e=1.5\times 10^{-3}$.}
	\label{fig:energy_norm_q_0}
\end{figure}

Furthermore, we see that $\lVert V_{\epsilon}\rVert_E$ depend on two parameters $\epsilon$ and $c_0$, for a fixed regular parameter $q$ of the Hayward black hole. Therefore, we can show contour plots in the two-dimensional parameter plane ($c_0,\epsilon$). For example, Fig.\ref{fig:energy_norm_q_0} is a contour plot with $q=0$. In Fig.\ref{fig:energy_norm_q_0}, the norm of $V_{\epsilon}$ on each contour line is equal. Then, we take parameters along one of the contour lines. These processes constitute the so-called equal energy method. It should be noted that, in terms of displaying results (see Fig.\ref{fig:G_factor_equal_energy} and Fig.\ref{fig:H_factor}), we still use $c_0$ as the horizontal axis and the results of the $\mathcal{G}$ factor or the $\mathcal{H}$-factor as the vertical axis.

\begin{figure}[htbp]
	\centering
	\includegraphics[width=0.45\textwidth]{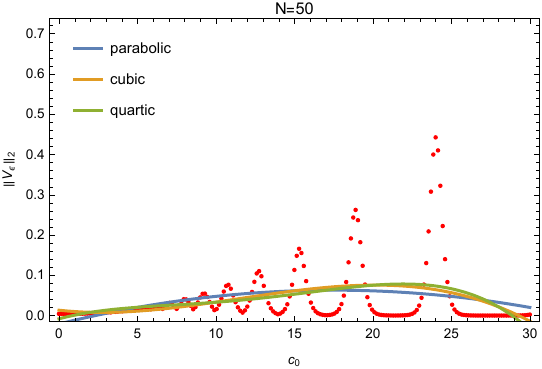}
    \hspace{0.5cm}
    \includegraphics[width=0.45\textwidth]{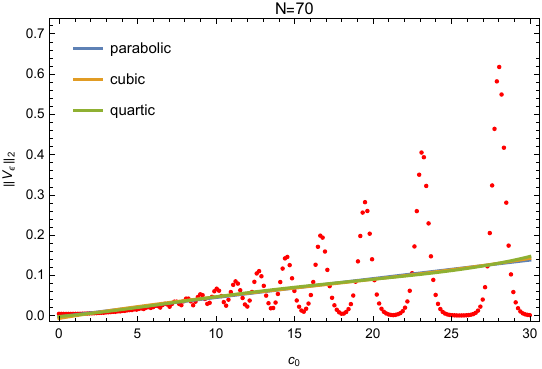}
    \includegraphics[width=0.45\textwidth]{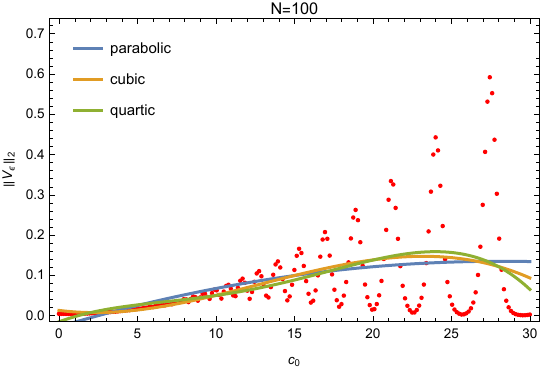}
     \hspace{0.5cm}
    \includegraphics[width=0.45\textwidth]{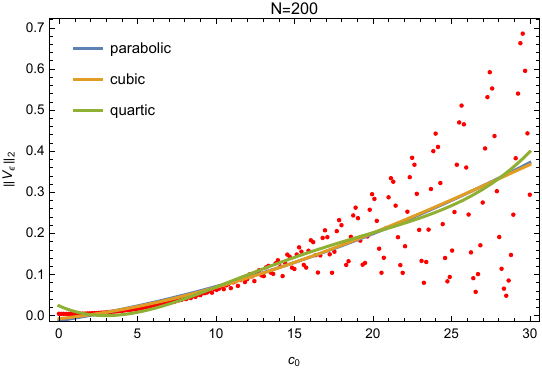}
    \includegraphics[width=0.45\textwidth]{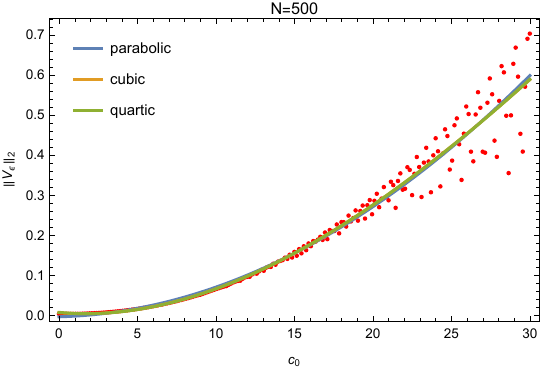}
    \hspace{0.5cm}
    \includegraphics[width=0.45\textwidth]{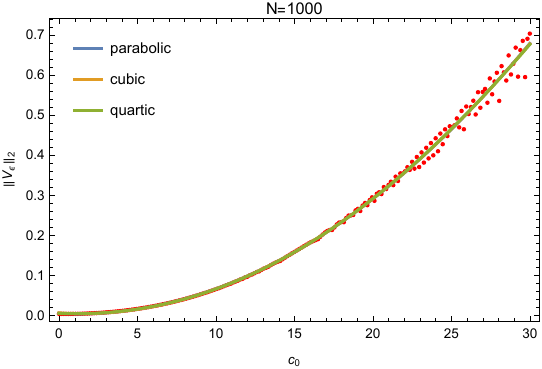}
    \includegraphics[width=0.45\textwidth]{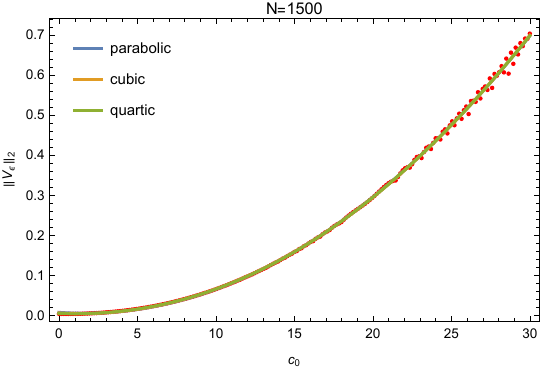}
     \hspace{0.5cm}
    \includegraphics[width=0.45\textwidth]{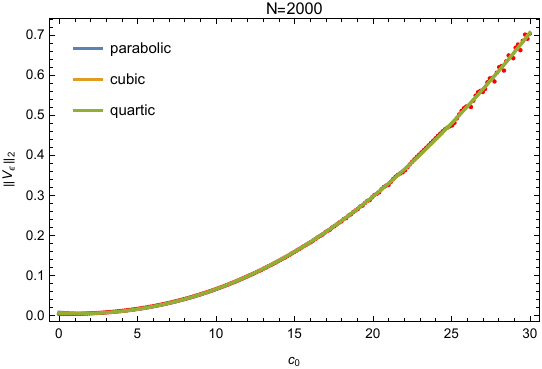}
	\caption{The $2$-norm of $V_\epsilon$ as a function of the location $c_0$ for $l=2$, $\epsilon=10^{-3}$, in units $r_{+}=1$. The plot range of $c_0$ is $[0,30]$, and the interval between adjacent red points is given by $\Delta c_0=0.15$. For each panels, we fit these points using quadratic, cubic, and quartic polynomials, where the blue line represents the quadratic polynomial, the yellow line represents the cubic polynomial, and the green line represents the quartic polynomial.}
	\label{fig:convergent_test}
\end{figure}

\begin{figure}[htbp]
	\centering
	\includegraphics[width=0.45\textwidth]{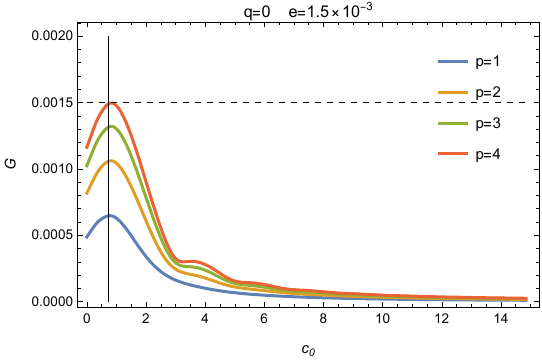}
    \hspace{0.5cm}
    \includegraphics[width=0.45\textwidth]{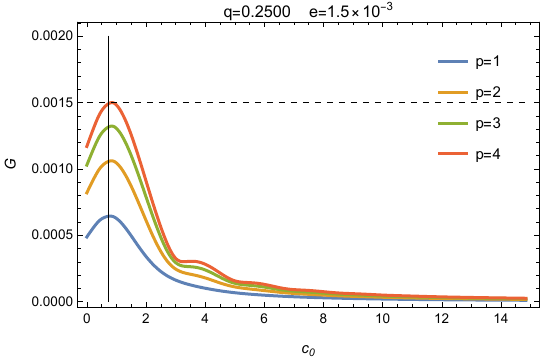}
     \includegraphics[width=0.45\textwidth]{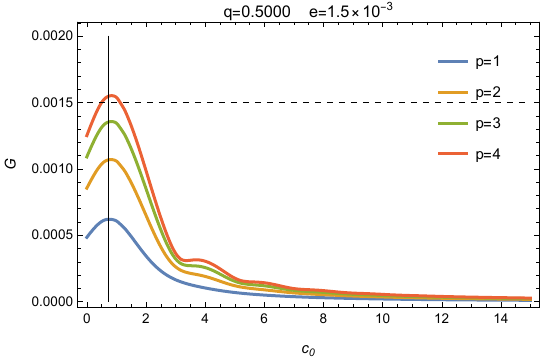}
     \hspace{0.5cm}
    \includegraphics[width=0.45\textwidth]{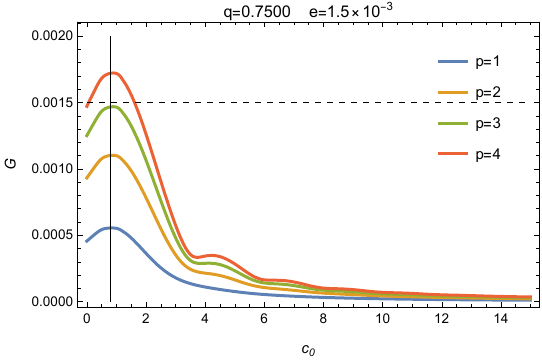}
	\caption{For the perturbation energy $e=1.5\times 10^{-3}$, the $\mathcal{G}$-factor varies with the position of the bump $c_0$ in terms of different $q$ and $p$. Each panel describes the variation of the $\mathcal{G}$-factor changing with $c_0$ and $p$ at a fixed $q$. Here, the vertical black line in each panel stands for the location of the peak of original potential for different $q$. The horizontal dashed line represents the energy norm of the bump.}
	\label{fig:G_factor_equal_energy}
\end{figure}

\begin{figure}[htbp]
	\centering
	\includegraphics[width=0.45\textwidth]{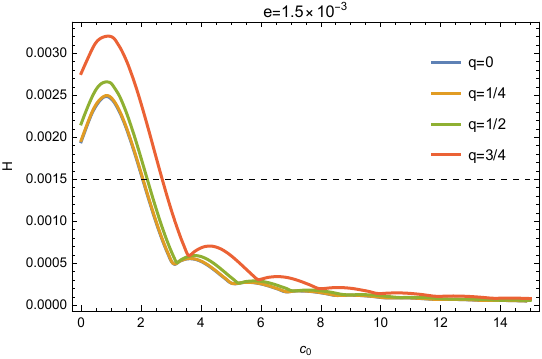}
    \hspace{0.5cm}
    \includegraphics[width=0.45\textwidth]{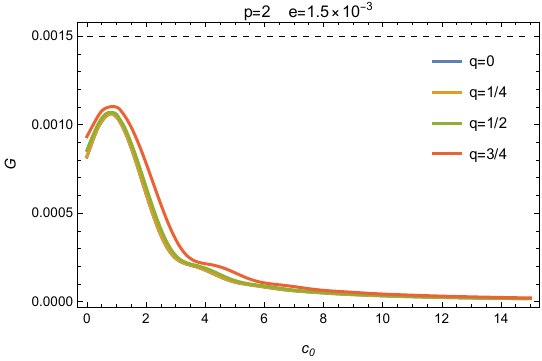}
	\caption{For the perturbation energy $e=1.5\times 10^{-3}$, the $\mathcal{H}$-factor varies with the position of the bump $c_0$ in terms of different $q$ for the left panel. For $p=2$ in Eq.(\ref{Gfactor}), the $\mathcal{G}$-factor varies with the position of the bump $c_0$ in terms of different $q$ for the right panel.}
	\label{fig:H_factor}
\end{figure}

Motivated by the recent work~\cite{Boyanov:2024fgc}, before computing the norm of the bump $V_\epsilon$, we are supposed to consider the convergence of such norm in terms of the resolution $N$. Without loss of generality, and considering the importance of the Schwarzschild black hole, we will do the convergent test of the norm of $\lVert V_\epsilon\rVert$ within $q=0$. In fact, different norms do not qualitatively alter the convergence of the norm of $\lVert V_\epsilon\rVert$~\cite{Boyanov:2022ark}. Therefore, we take the norm $\lVert\cdot\rVert$ into the standard $2$-norm as a matter of convenience and use the Chebyshev-Labatto grid to complete the discretization. In other words, $W$ in Eq.(\ref{Veps_norm}) is the identity matrix avoiding the time-consuming of the Cholesky decomposition of Gram matrix and the matrix multiplication when $N$ is very large. In addition, we take $l=2$, $r_{+}=1$ and $\epsilon=10^{-3}$ as the necessary test parameters. Within different resolutions $N$, the results of the convergent test of the norm of $\lVert V_\epsilon\rVert$ are shown in Fig.\ref{fig:convergent_test}.

As the Fig.\ref{fig:convergent_test} shows, for the situation of the low resolution, as $c_0$ increases, the norm of $V_\epsilon$ will periodically enter a valley. At the same time, we can clearly see that the set of these red points has a clear envelope at low resolution. In scenarios with relatively low resolution, when $c_0$ is large, the periodic occurrence of valleys is attributed to the fact that the width of the bump (\ref{perturbed_effective_potential}) in $\sigma$ coordinate is too narrow, resulting in too few grid points within the location of its support set. As the resolution gradually increases, these valleys will gradually disappear. Further, we find that as the resolution increases, the fitted curve tends to stabilize and gradually becomes the envelope curve in the low resolution case. Based on the above discussion, we have sufficient reason to believe that as $c_0$ increases by remaining the amplitude unchanged, the energy norm of the bump also increases. From the perspective of pseudospectrum, the increase of the norm is the reason for the significant migration of the QNM frequencies~\cite{Cheung:2021bol}.

Considering that improving the resolution will significantly increase the consumption of time, yet the resolution cannot be too low, otherwise the convergence of the norm will become a catastrophic issue. Thus, in our study, we adopt a compromise in selecting the resolution and the maximum value of $c_0$, with $N=500$ and $c_{0\text{max}}=15$. These choices are sufficient to reflect the stability of the greybody factor of the Hayward black hole within the equal energy method. For $q=0$ as a typical example, the energy norm contour lines of the bump are shown in Fig.\ref{fig:energy_norm_q_0} ,where $\epsilon_{\text{min}}=0$, $\epsilon_{\text{max}}=1/1000$, $c_{0\text{min}}=0$ and $c_{0\text{max}}=15$ are considered.

Under the equal energy condition ($e=1.5\times 10^{-3}$), we present the behaviors of the $\mathcal{G}$-factor and $\mathcal{H}$-factor as the location of bump varies in Fig.\ref{fig:G_factor_equal_energy} and Fig.\ref{fig:H_factor}. As these figures shown, we can easily observe some features that are similar to the equal amplitude case: First, both the $\mathcal{G}$-factor and $\mathcal{H}$-factor still reach their maximum when the bump is located at the peak of the original effective potential, the $\mathcal{G}$-factor and $\mathcal{H}$-factor are still bounded by the order of magnitude of the perturbation energy $\mathcal{O}(e)$. Second, $\mathcal{G}$-factor and $\mathcal{H}$-factor are also insensitive to the regular parameter $q$ of Hayward black hole. So we can conclude that in the case of the equal energy, the greybody factor is still stable. 

However, the significant difference between the equal amplitude and equal energy method lies in their distinct asymptotic behavior as the location of the bump increases. Specifically, for equal amplitude case, the $\mathcal{G}$-factor and $\mathcal{H}$-factor approach a specific value as $c_{0}$ increases, where such specific value for the $\mathcal{G}$-factor depends on the choice of $p$. In contrast, for equal energy case, the $\mathcal{G}$-factor and $\mathcal{H}$-factor converge to zero as $c_{0}$ increases. This is due to the fact that, under condition of a constant energy for a bump, as $c_0$ increases, the amplitude diminishes accordingly. Accordingly, a small perturbation at infinity will have almost no effect on the greybody factors, in which the size of the perturbation is measured by the energy (norm).

%======================================%
%<<<<<<<<<<< Conclusions >>>>>>>>>>>>>%
%======================================%
\section{conclusions and discussion}\label{sec: conclusions}
In this work, we have investigated the stability of the greybody factor of Hayward black hole, and we particularly concentrate on the massless scalar field propagating in the Hayward black hole. First, we find that for non-extreme Hayward black holes, the construction of hyperboloidal framework is completely feasible, which will be used in the equal energy method. The construction of height function for the compact coordinate transformations (\ref{hyperboloidal_coordinate_transformation}) is described in Appendix.\ref{hyperboloidal_coordinate}. To further verify the validity of the obtained hyperboloidal coordinates, we analyze the characteristic speeds of the hyperbolic equation (\ref{regular_equation}). As a result, the outgoing characteristic speed vanishes on the event horizon and the ingoing characteristic speed vanishes on the null infinity. This tells us that the boundary conditions of QNMs will be automatically embedded in Eq.(\ref{regular_equation}). The equation (\ref{regular_equation}) can be rewritten as a matrix differential equation, and we solve the eigenvalue problem to obtain the QNM frequencies (see Table \ref{QNMs}) as a byproduct of the hyperboloidal approach. The results are  consistent with Refs.\cite{Flachi:2012nv,Toshmatov:2015wga}. 

In Sec.\ref{sec: greybody}, we have studied the stability of the greybody factor against a small bump correction of effective potential. Our investigation delves into the effects that various parameters exert on the stability of the greybody factor associated with Hayward black holes within two different parameter-choosing contexts, i.e., the equal amplitude method and the equal energy method. To quantitatively describe stabilities, we introduce two factors, i.e. $\mathcal{G}$-factor and $\mathcal{H}$-factor, where their definitions can be found in Eq.(\ref{Gfactor}) and Eqs.(\ref{Hfactor1})-(\ref{Hfactor2}). For the equal amplitude method, the results are shown in Fig.\ref{fig:G_factor_equal_eps} and Fig.\ref{fig:H_factor_equal_eps} associated with different $q$. When considering a bump with equal amplitude $\epsilon$, results of two factors indicate that small perturbations do not lead to significant changes of the greybody factors.

Utilizing the hyperboloidal framework constructed in Sec.\ref{sec:The hyperboloidal framework}, we are able to physically delineate the energy emanating from the bump and provide equal energy curves in the parameter space determined by two parameters ($\epsilon$, $c_{0}$). Subsequently, we can study the changes in the greybody factor in the condition of equal energy. Then we find that, when the energy carried by the bump is small, the greybody factor remains stable, for the $\mathcal{G}$-factor and the $\mathcal{H}$-factor staying at very low values. And both methods of characterizing stability show that the greybody factor of the Hayward black hole is a stable physical quantity~\cite{Rosato:2024arw}.

For these two different parameters selection methods, our calculations reveal that the greybody factor is stable, with the maximum values of the $\mathcal{G}$-factor and $\mathcal{H}$-factor occurring when the bump is located at the peak of the original effective potential. These two factors have the same order of magnitude as the bump's amplitude $\epsilon$ or energy $e$. However, as the bump's location gradually increases, the asymptotic behavior differs significantly between equal amplitude method and equal energy method. For the equal amplitude case, the $\mathcal{G}$-factor and $\mathcal{H}$-factor approach a particular value, but for equal energy case, both the $\mathcal{G}$-factor and $\mathcal{H}$-factor converge to zero. It is foreseeable that, at infinity, small perturbation will have almost no effect on the greybody factor of a Hayward black hole with the equal energy method. Given that small perturbations at infinity should have no significant impact on the entire physical system. Thus, it may be more reasonable to consider perturbations with equal energy not equal amplitude, because when fixing the bump's amplitude, if the location of the bump is very far from the black hole, the energy corresponding to the perturbation will be large (see Fig.\ref{fig:convergent_test}), causing a non-negligible gravitational effect.

As a supplement, we provide an explanation that the $\mathcal{G}$-factor and the $\mathcal{H}$-factor are insensitive to parameter $q$. It is known that the greybody factor is entirely determined by the effective potential. Therefore, the impact of the regular parameter $q$ on the $\mathcal{G}$-factor and the $\mathcal{H}$-factor depends on how $q$ influences the effective potential $V_{l}^{\epsilon}(r)$. In order to clearly demonstrate the effect of $q$ on the effective potential, we expand $V_{l}^{\epsilon}(r)$ with $r_{+}=1$ as a series in terms of $q$, i.e.,
\begin{eqnarray}\label{V_expansion}
    V_{l}^{\epsilon}(r)&=&\frac{(r-1)(l^2 r+l r+1)}{r^4}+4\epsilon\text{sech}\Big[2\Big(r+\ln(r-1)-c_0 \Big)\Big]^2\nonumber\\
    &&+\frac{q^4}{r^7}\Bigg[5+(l^2+l-4) r-2 r^3-(l^2+l-1)r^4+8 r^6 \epsilon\text{sech}\Big[2 \Big(r+\ln(r-1)-c_0 \Big)\Big]^2\nonumber\\
    &&\times\Bigg(\tanh \Big[2 \Big(r+\ln(r-1)-c_0 \Big)\Big]\times \Big(-2 c_0 r+2 r^2-r-2+4r\ln\Big(\frac{r}{r-1}\Big)\Big)-r\Bigg)\Bigg]\nonumber\\
    &&+\mathcal{O}(q^{6})\, .
\end{eqnarray}
The first term of Eq.(\ref{V_expansion}) is nothing but the potential in the Schwarzschild black hole. From Eq.(\ref{V_expansion}), we can find that the effect of $q$ on the effective potential only manifests itself in higher-order terms. The contribution of the lowest order is $\mathcal{O}(q^4)$. Since we have $q\in[0,1]$, the variation of $q$ causes only a small change in the effective potential $V_{l}^{\epsilon}(r)$. Furthermore, it is known that there exists a relationship between the black hole's greybody factors and QNMs, which is given by Eq.(3.5) in~\cite{Konoplya:2024lir}. From such a relationship, the greybody factor is mainly determined by the fundamental mode. A lot of research on the stability of black hole's QNMs indicates that the fundamental mode is stable under perturbations, while instability mainly occurs in higher-order overtones. So after adding the bump, the change of the fundamental mode is not significant. According to Appendix.\ref{app: QNMs}, the fundamental mode ($n=0$) of the Hayward black hole is insensitive to variations in the regular parameter $q$. Combining it with the relationship between the greybody factor and QNMs, we can infer that the $\mathcal{G}$-factor and the $\mathcal{H}$-factor, which characterize the stability of the greybody factor, are also insensitive to the regular parameter $q$.

The greybody factor is a fundamental concept in black hole physics and gravitational wave research, with significant applications in understanding radiation (such as Hawking radiation) and gravitational wave. Hawking radiation is the thermal radiation emitted due to quantum effects near the event horizon of a black hole. However, because of the black hole's gravitational barrier, some of the radiation is suppressed or absorbed. The greybody factor quantifies the ability of radiation to overcome this potential barrier and escape to infinity, determining the probability of different frequencies or energies of Hawking radiation escaping the black hole. Specifically, the greybody factor helps us to understand the differences in radiation emitted by various types of black holes, such as charged black holes and regular black holes. From the greybody factor, one can get the power spectra for Hawking radiation~\cite{Zhang:2020qam,Hawking:1975vcx}:
    \begin{eqnarray}\label{Hawking_radiation}
        \frac{\mathrm{d}^2 E}{\mathrm{d}t\mathrm{d}\omega}=\frac{1}{2 \pi} \sum_l \frac{N_l\left|\Gamma_{l}(\omega)\right|^2 \omega}{e^{\omega / T}-1}\, ,\quad N_l=\frac{(2 l+d-3)(l+d-4)!}{l!(d-3)!}\, ,
    \end{eqnarray}
where $T$ is the black hole's temperature. Here, $l$ refer to different angular modes, and $d$ is the dimension of the spacetime. The study of the stability of the greybody factor also inspires us to investigate the stability of the Hawking radiation power spectrum.

%======================================%
%<<<<<<<<<< Acknowledgement >>>>>>>>>>>%
%======================================%
\section*{Acknowledgement}
We are grateful to Li-Ming Cao and Yu-Sen Zhou for helpful discussions. This work was supported by the National Natural Science Foundation of China with grants No.12235019 and No.11821505.

\appendix
\section{The hyperboloidal framework for Hayward black hole}\label{hyperboloidal_coordinate}
In this appendix, we show the following process to give the hyperboloidal coordinate of Hayward black hole~\cite{PanossoMacedo:2023qzp}. Specifically, the coordinate transformations are given by
\begin{eqnarray}\label{hyperboloidal_coordinate_transformation}
    t=r_{+}(\tau-H(\sigma))\, ,\quad r=\frac{r_{+}}{\sigma}\, , 
\end{eqnarray}
where $r_{+}$ is the event horizon. In terms of $\sigma$, the metric function can be expressed as
\begin{eqnarray}
    \mathcal{F}(\sigma)=f(r(\sigma))=\frac{(1-\sigma)(1-q^2\sigma)[1+q^2(1+\sigma)]}{1+q^2+q^4\sigma^3}\, .
\end{eqnarray}
The derivative of the dimensionless tortoise coordinate $x(\sigma)$ is expanded as
\begin{eqnarray}
    x^{\prime}(\sigma)=-\frac{1}{\sigma^2\mathcal{F}(\sigma)}=-\frac{1}{\sigma^2}+\Big(-1 - q^2 + \frac{q^2}{1 + q^2}\Big)\frac{1}{\sigma}+\mathcal{O}(1)\, ,\quad \text{as}\quad \sigma\to0\, .
\end{eqnarray}
The leading terms contribute with the singular quantities
\begin{eqnarray}
    x_0(\sigma)=\frac{1}{\sigma}+\Big(-1 - q^2 + \frac{q^2}{1 + q^2}\Big) \ln\sigma\, .
\end{eqnarray}

At the event horizon $\sigma=1$, the integration around the horizon $\sigma=1$ yields
\begin{eqnarray}
    x_{+}(\sigma)=\frac{1}{K_{+}(1)}\ln|\sigma-1|\, ,
\end{eqnarray}
where
\begin{eqnarray}\label{K_p}
    K_{+}(1)=\frac{(1 - q^2) (1 + 2 q^2)}{1 + q^2 + q^4}\, .
\end{eqnarray}
At the inner horizon $\sigma=1/q^2$, the integration around the horizon $\sigma=1$ yields
\begin{eqnarray}
    x_{-}(\sigma)=\frac{q^2}{K_{-}(1/q^2)}\ln\left|\sigma-\frac{1}{q^2}\right|\, ,
\end{eqnarray}
where
\begin{eqnarray}\label{K_m}
    K_{-}(1/q^2)=\frac{-2 + q^2 + q^4}{1 + q^2 + q^4}\, .
\end{eqnarray}
The regular piece $x_{\text{reg}}(\sigma)$ is satisfied with
\begin{eqnarray}
    x^{\prime}_{\text{reg}}(\sigma)=x^{\prime}(\sigma)-x^{\prime}_{+}(\sigma)-x^{\prime}_{-}(\sigma)-x^{\prime}_0(\sigma)\, .
\end{eqnarray}

Therefore, derivatives of two height functions $H^{\text{in-out}}(\sigma)$ and $H^{\text{out-in}}(\sigma)$ associated with $\sigma$ are 
\begin{eqnarray}
    \frac{\mathrm{d}H^{\text{in-out}}(\sigma)}{\mathrm{d}\sigma}=-\frac{1}{\sigma^2\mathcal{F}(\sigma)}+\frac{2}{\sigma^2}-\Big(-2 - 2q^2 + \frac{2q^2}{1 + q^2}\Big)\frac{1}{\sigma}\, ,
\end{eqnarray}
and
\begin{eqnarray}
    \frac{\mathrm{d}H^{\text{out-in}}(\sigma)}{\mathrm{d}\sigma}=\frac{1}{K_{+}(1)}\frac{2}{\sigma-1}+\frac{q^2}{K_{-}(1/q^2)}\frac{2}{\sigma-{1}/{q^2}}+\frac{1}{\sigma^2\mathcal{F}(\sigma)}\, .
\end{eqnarray}
Note that when $q=0$, i.e., the case of the Schwarzschild black hole, two height functions are equal.

\section{The quasinormal modes}\label{app: QNMs}

In this appendix, we show QNM frequencies of Hayward black hole by using the pseudo-spectral method. Note that for the QNM frequencies with $n\leq3$, it is accurate enough to solve them with the resolution $N=100$.

\begin{table}[ht]
 \renewcommand{\arraystretch}{1.5}
\begin{tabular}{llcccc}
 \hline
 
$l$ ~~~~~~& $n$ ~~~~~~& $q=0$~~~& $q=1/4$ ~~~& $q=1/2$ ~~~& $q=3/4$\cr

 \hline
0  & 0 & $0.1105+\imath~0.1049$ ~~~& $0.1107+\imath~0.1046$ ~~~& $0.1130+\imath~0.1011$ ~~~& $0.1144+\imath~0.0906$\\
  & 1 & $0.0861+\imath~0.3480$ ~~~& $0.0864+\imath~0.3469$ ~~~& $0.0876+\imath~0.3325$ ~~~& $0.0507+\imath~0.3196$\\
  & 2 & $0.0755+\imath~0.6010$ ~~~& $0.7560+\imath~0.5989$ ~~~& $0.0709+\imath~0.5748$ ~~~& $0.0427+\imath~0.5578$\\
   & 3 & $0.0704+\imath~0.8549$ ~~~& $0.0696+\imath~0.8521$ ~~~& $0.0569+\imath~0.8173$ ~~~& $0.0392+\imath~0.8260$ \\
\hline
1   & 0 & $0.2929+\imath~0.0977$ ~~~& $0.2933+\imath~0.0974$ ~~~& $0.2973+\imath~0.0946$ ~~~& $0.3048+\imath~0.0868$\\
   & 1 & $0.2644+\imath~0.3063$ ~~~& $0.2650+\imath~0.3054$ ~~~& $0.2709+\imath~0.2948$ ~~~& $0.2756+\imath~0.2672$ \\
   & 2 & $0.2295+\imath~0.5401$ ~~~& $0.2302+\imath~0.5384$ ~~~& $0.2362+\imath~0.5167$ ~~~& $0.2220+\imath~0.4677$\\
   & 3 & $0.2033+\imath~0.7883$ ~~~& $0.2039+\imath~0.7856$ ~~~& $0.2063+\imath~0.7519$ ~~~& $0.1642+\imath~0.7008$\\
\hline
2   & 0 & $0.4836+\imath~0.0968$ ~~~& $0.4842+\imath~0.0965$ ~~~& $0.4904+\imath~0.0938$ ~~~& $0.5035+\imath~0.0865$\\
   & 1 & $0.4639+\imath~0.2956$ ~~~& $0.4645+\imath~0.2949$ ~~~& $0.4722+\imath~0.2860$ ~~~& $0.4850+\imath~0.2620$ \\
   & 2 & $0.4305+\imath~0.5086$ ~~~& $0.4315+\imath~0.5072$ ~~~& $0.4410+\imath~0.4899$ ~~~& $0.4495+\imath~0.4453$ \\
   & 3 & $0.3939+\imath~0.7381$ ~~~& $0.3949+\imath~0.7358$ ~~~& $0.4051+\imath~0.7081$ ~~~& $0.3993+\imath~0.6408$ \\
 \hline
\end{tabular}
\caption{QNM frequencies for massless scalar perturbation for $q=0$, $1/4$, $1/2$, $3/4$ and $M=1$ for the Hayward black hole. These results are from the resolution $N=100$.}
\label{QNMs}
\end{table}

\bibliography{reference}
\bibliographystyle{apsrev4-1}

\end{document}